\documentclass[a4paper,twocolumn]{article}
\usepackage{lineno,hyperref}
\usepackage[T1]{fontenc}
\usepackage{graphicx,amsmath}
\usepackage{soul,color}
\title{Fluorine patterning of graphene: Effects of fluorine content and temperature}

\author{Ruslan D. Yamaletdinov$^{*,1,2}$, Yaroslav A. Nikiforov$^{3}$,\\ Lyubov G. Bulusheva$^{2}$, Alexander V. Okotrub$^{2}$}
\date{
	$^1$ Boreskov Institute of Catalysis SB RAS, 5,  Lavrentiev Ave., 630090  Novosibirsk, Russia\\
	$^2$ Nikolaev Institute of Inorganic Chemistry	SB RAS, 3, Lavrentiev Ave., 630090, Novosibirsk, Russia\\
	$^3$ Novosibirsk State University, 1, Pirogova str., 630090, Novosibirsk, Russia\\
	$^*$ \textit{yamaletdinov@niic.nsc.ru}\\
	\today}

\begin{document}
	\setcounter{secnumdepth}{0}
	\twocolumn[
  \begin{@twocolumnfalse}
    \maketitle
    \begin{abstract}
      In this paper we present a successful approach for the generation of partially fluorinated graphene structures. A computationally simple model optimized on a large DFT dataset quickly and precisely  predicts experimentally observed structures. From the analysis of the structural diversity of fluorinated graphene in a wide range of synthesis temperatures, the general structural patterns are identified and the conditions for their achievement are determined. In addition, to facilitate further studies of fluorinated graphene, we present a ready-to-use GenCF code that implements the described structure generator.
    \end{abstract}
  \end{@twocolumnfalse}
  ]
	
\section*{Introduction}
	There are three most studied derivatives of chemically functionalized graphene: graphene oxide, hydrogenated graphene and fluorinated graphene~\cite{Mkhoyan2009,Sofo2007,Nair2010,Craciun2013}. In each of them adatoms turn sp$^2$ carbon to sp$^3$ one upon $\sigma$ bonding. In graphene oxide, atoms are attached randomly or form distinct oxidized regions mainly covered by epoxy and hydroxyl groups without any specific patterns~\cite{Mkhoyan2009,Erickson2010}. The hydrogenated graphene (HG) and fluorinated graphene (FG) have similar structural features: particularly, the neigboring adatoms prefer to attach to different sublattice positions at opposite sides of the graphene sheet. This preference makes a chair conformation and adatom pairs the most stable structures for fully and partially functionalized graphene, respectively~\cite{Sofo2007,Nair2010,Paupitz2013,Johns2013,Boukhvalov2008,Zhou2014}.
 A serious difference between FG and HG is the sign of the binding enthalpy of a single adatom with respect to the gas~\cite{ribas2011patterning}. 
	
	For our purposes, we classified the synthesis methods of  FG~\cite{Feng2016} into: substitution of oxygen groups in graphene oxide, exfoliation of fluorinated graphite and direct fluorination of a graphene or activated graphitic materials. While the structure of the resulting FG is determined by the structure of the precursor in the case of the graphene oxide and fluorinated graphite, description of the structures formed via direct fluorination is being discussed. Direct fluorination can be carried out using a liquid (e.g. BrF$_3$~\cite{Asanov2013}, HF~\cite{Nebogatikova2015}) or gaseous (e.g. F$_2$~\cite{Cheng2010}, XeF$_2$~\cite{Jeon2011}) fluorinating agent or by plasma treatment (e.g. CF$_4$ and SF$_6$ plasma~\cite{Struzzi2017}). The direct method for studying the structure of the obtained samples is atomic resolution microscopy. Unfortunately, this characterization is relatively rare in use due to the complexity of sample preparation, and measurement issues, and the lack of need for such characterization in many cases. Widely used methods for the qualitative description of the FG structure include X-ray photoelectron spectroscopy (XPS), allowing the determination of various chemical forms~\cite{Feng2016}, Raman spectroscopy, showing the graphene lattice disordering~\cite{Ferrari2007}, and nuclear magnetic resonance (NMR), characterizing both the number of different forms and their relative positions~\cite{Vyalikh2013}.
	
   Theoretical approaches to study FG structures are density functional theory (DFT) and molecular dynamics (MD) simulations. DFT calculations allow to accurately determine the energy of specific structures, such as various conformations of fully FG~\cite{Samarakoon2011}, fluorinated or bare carbon chains, roads and quantum dots~\cite{ribas2011patterning}, specific pairs and triples and their impact on the surroundings~\cite{Zhou2014,ribas2011patterning}, evenly distributed fluorine~\cite{Santos2014}, periodic patterns~\cite{Boukhvalov2016} and manually set patterns~\cite{Vyalikh2013,Langer2019,Makarova2017}. Such information  gives opportunity to study the thermodynamically most stable structures and can help to establish fluorination mechanisms, but it can hardly predict structural variations under different synthesis conditions. It is generally accepted that MD simulations are the most optimal tool for these purposes. In several works, fluorination modeling or thermal properties calculations of FG were carried out using ReaxFF force field (e.g. Refs.~\cite{Paupitz2013,Singh2013,Yamaletdinov2020}). Due to the limitations of ReaxFF on accounting for the specific influence of distant neighbors~\cite{VanDuin2001}, this approach is limitedly applicable for correct simulations of conjugated systems such as low and medium fluorinated graphene. 
   
	In this paper, generating FG structures where the parameters were optimized using a large DFT dataset. We show that adjustable distribution generates the structures corresponding to different synthesis conditions and allows us to analyze the distribution of patterns. 
\section{Simulation Details}
\begin{figure}[!h]
		\begin{center}
			\includegraphics[width=0.5\columnwidth]{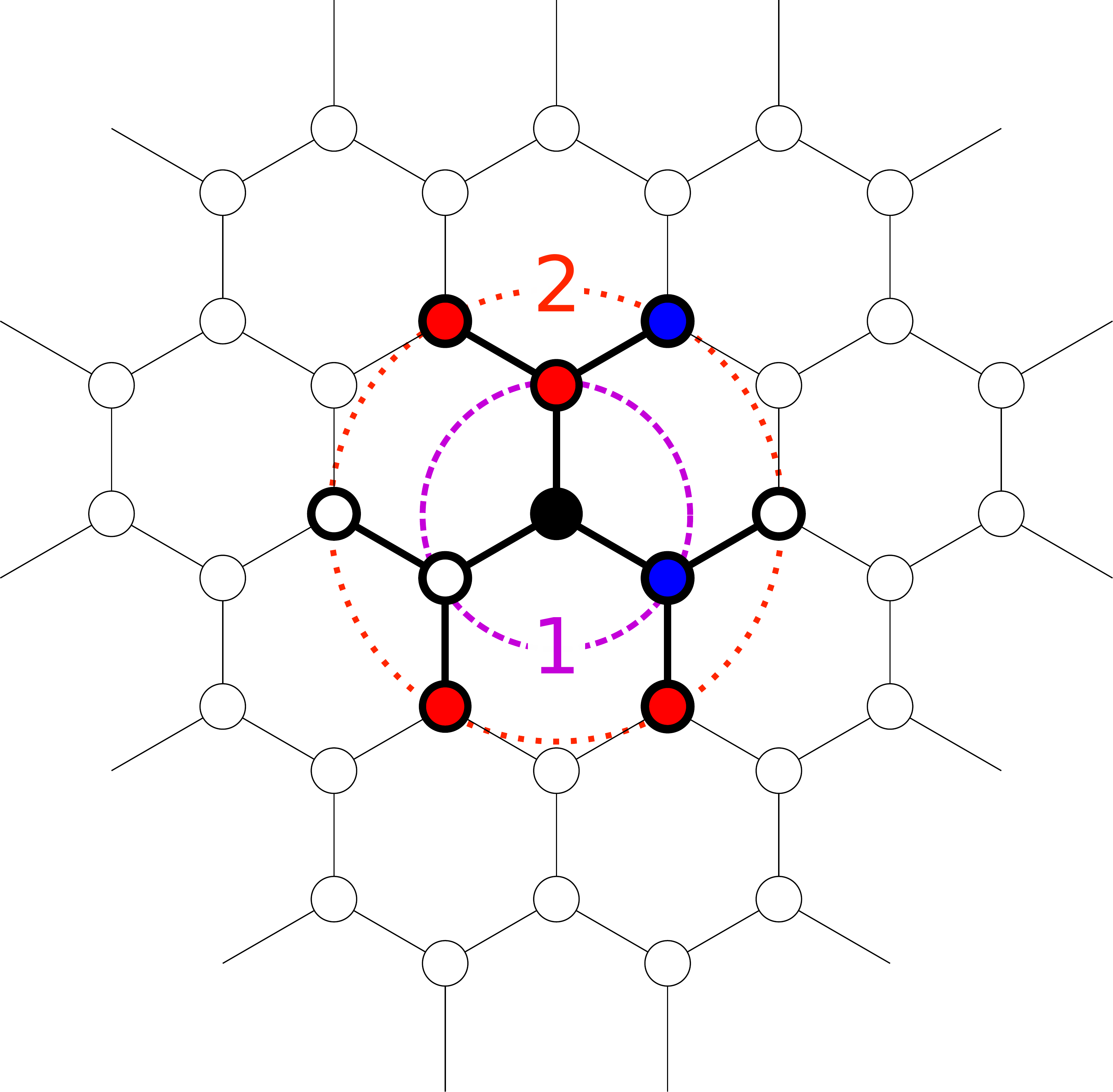}
		\end{center}
		\caption{An example of fluorine atoms distribution in a graphene lattice. Bold lines depict the elementary structure. Black atom is the central atom (see description under Eq.~\ref{eq:prob}). Purple and orange dashed circles correspond to first and second coordination spheres. Red and blue colored atoms correspond to up and down oriented fluorine, respectively. }
		\label{str_examp}
	\end{figure}

	DFT calculations were carried out in Jaguar software ver. 10.3~\cite{Jaguar} within the B3LYP exchange-correlation functional~\cite{becke1988density,Lee1988,Vosko1980,slater1974quantum} along with 6-31G basis sets~\cite{Krishnan1980} for all atoms. In every single calculation a hydrogen-terminated graphene flake of a C$_{80}$H$_{22}$ composition was used. Since we were interested in the surrounding influence on fluorine binding energy, we focused on a small elementary structure with 10 atoms, which includes central atom and first two coordination spheres (Fig.~\ref{str_examp}). From 1 to 10 fluorine atoms were randomly scattered over the elementary structure. To avoid the repetition of fluorine distribution, the symmetry operations of the D$_{3h}$ point group were applied to the carbon  skeleton of the elementary structure.  Geometry of the obtained 500 differnt FG flakes was optimized using DFT. We understand that for such conjugated systems the influence of third surrounding sphere might be appreciable, while the strongest C-F bonding occurs in the case of orthofluorination~\cite{ribas2011patterning}.  Also, addition of 9 more sites for fluorine atoms would greatly complicate the combinatory task.
	
	DFT results were further compared with ReaxFF MD energies. MD simulations were carried out in LAMMPS~\cite{Plimpton1995} package with ReaxFF force field~\cite{Aktulga2012} optimized for fluorographene CF~\cite{Singh2013}. For a single fluorine atom the 2.9 eV C-F binding energy was obtained. This value is comparable with that for fluorographene CF, although it should be much less~\cite{Santos2014}. The 2.9-eV value is extremely overestimated in relation to DFT values ($\times2-4$ times, average $\times2.5$ times for our low-fluorinated elementary structures).  We also ensure, that our DFT results are consistent with other works (e.g.~\cite{Zhou2014,ribas2011patterning,Langer2019}) at least for graphenes with a low fluorination degree. The overestimation of C-F bonding and possibility of direct accounting of only the nearest neighbors is drawbacks of the ReaxFF method for conjugate systems. We warn to be careful when using ReaxFF FC for thermodynamic calculations of non-fluoroalkane systems.
	
\begin{figure}[!h]
		
		\includegraphics[width=\columnwidth]{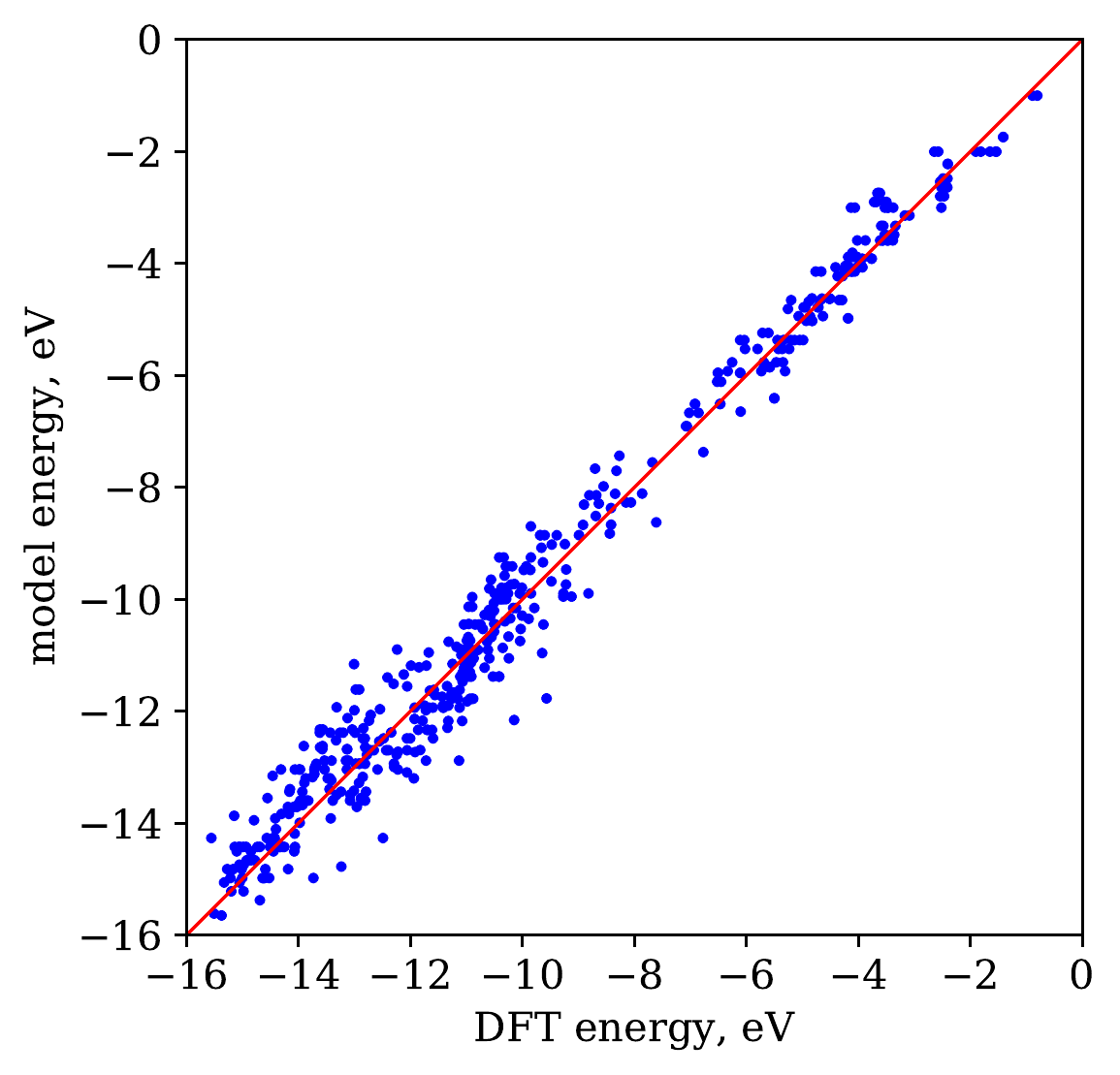}
		\caption{Binding energies of fluorine atoms to graphene flake calculated using the model desctibed by Eq.~\ref{eq_E} vs DFT data for various patterns in the elementary structure. Red line is a guide to eye.}
		\label{fig_E}
	\end{figure}

	To predict fluorine binding energies we tried to find any hard correlations between obtained geometries and the corresponding energies within a few regression methods (like principal component or support vector regressions). Unfortunately, this approach did not allow us to achieve satisfying results. It was reported that a simple short-range pair potential in Ising-like interaction model exhibits a good prognostic potential for HG~\cite{Gargiulo2014}. We have also conclude that such approach could describe FG system. In our model, to calculate the  binding energy of fluorine atoms to a graphene structure we use following equation:
	\begin{multline}
	E=E_f \left\Vert g\right\Vert^2+ \sum_{[i<j]} [a_1 g_i g_j + b_1 g_i^2 g_j^2]+\\
	\sum_{[[i<j]]} [a_2 g_i g_j + b_2 g_i^2 g_j^2],
	\label{eq_E}
	\end{multline} 
	where $g$ is the structure vector with elements equal to 0 or 1 or -1, depending on the absence, or upward, or downward orientation of fluorine atoms, respectively. First term is the "pure" C-F binding energy in the absence of any neighbors. The second and third terms are sums over the first and second order neighbors, respectively. Each sum contains two terms. We assume that, the $a_{1,2}$ terms are mostly responsible for the geometric distortion of the lattice caused by covalent attachment of fluorine and they are dependent on the relative orientation of the $i$-th and $j$-th atoms, while the $b_{1,2}$ terms reveal the orientation-independent energy corrections for the neighbors from the first and second surrounding fluorine spheres. A good agreement of $E$ values from Eq.~\ref{eq_E} with the above DFT energies (see Fig.~\ref{fig_E}) was obtained with the parameters presented in table~\ref{table_params}.
	
	\begin{table}[!h]
		\caption{Fitted parameters (in eV) of the model described in Eq.~\ref{eq_E}}
		
		\begin{center}
			\begin{tabular}{|c|c|c|c|c|}
				\hline
				$E_f$ & $a_1$& $a_2$& $b_1$& $b_2$\\
				\hline
				-1.00 & 0.278& 0.079& -0.863& 0.181\\
				\hline
			\end{tabular}
			\label{table_params}
		\end{center}
		
	\end{table}

\section*{Results and Discussion}

\begin{figure}[!h]
		\begin{center}
			\includegraphics[width=0.7\columnwidth]{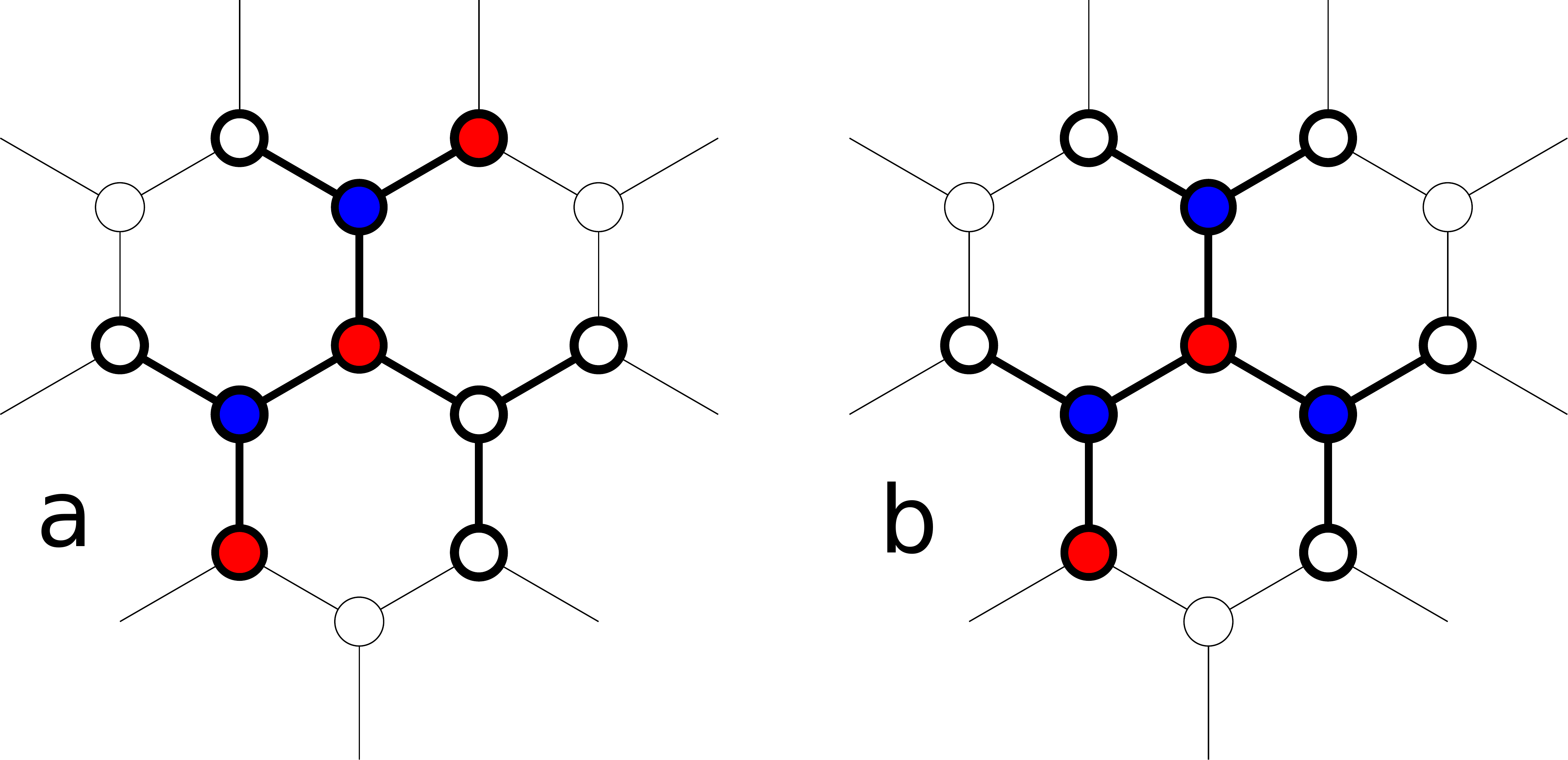}
			\caption{Two structures of C$_{10}$F$_5$ with the lowest energies. According to Eq.~\ref{eq_E}  $E_b-E_a=a_2+b_2$.} 
			\label{str_c}
		\end{center}
	\end{figure}

	The obtained results allow us to conclude that in a fluorination process, the growth of chains or island is energetically justified for alternating fluorination due to a low value of $b_1$. Ref.~~\cite{Johns2013} also mentioned that the binding energy of additional fluorine atoms increase. Orientation of the nearest neighbors has the greatest impact on the energy of FG and expectedly the lower $\sum_{[i<j]}g_i g_j$ the higher binding energy. It leads to an explicit predominance of pairs and chains over clusters of bound fluorine atoms in case of one-side fluorination. More interesting is the second neighbors influence. It was shown that in the case of only two fluorine atoms, the meta-position of the fluorine atoms in a carbon hexagon results in weaker binding even than single F atoms due to an increase in the number of unpaired localized electrons (for example see~\cite{Zhou2014,ribas2011patterning}). Since we used the set of random structures, our model can expand this statement: the presence of the second-order neighbors always suppresses the fluorine binding efficiency. To exemplify the consequence of this effect, the energies of the two most preferable fluorine patterns giving the C$_{10}$F$_5$ composition of  the elementary structure  are compared (Fig.~\ref{str_c}). Structure presented in Fig.~\ref{str_c} (a) is $E_b-E_a=a_2+b_2=0.26$ eV more stable than structure in Fig.~\ref{str_c} (b).  This can lead to the experimentally observed chain pattern of FG especially at a low synthesis temperature~\cite{Asanov2013,Bulusheva2002}. 
	
	In order to simulate the fluorination process, we propose a simple sequential scheme~\cite{Yamaletdinov2020}, where the probability of top ($p_i^+$) or bottom ($p_i^-$) fluorination of graphene at the $i$-th atom equals to:
	\begin{equation}
	p_i^\pm={\frac{\Omega_i^\pm \exp(-\beta E_i^\pm)}{\sum \Omega_j  \exp(-\beta E_j)}}, 
	\label{eq:prob}
	\end{equation}  
	where $E_i^\pm$ is the binding energy of fluorine with the $i$-th site, which is calculated as the difference between the energies of the structure (Eq.~\ref{eq_E}) with bound fluorine $E^\pm$ at the center of the elementary structure(see Fig.~\ref{str_examp}) and the primary structure without this central fluorine $E^0$: $E_i^\pm = E^\pm-E^0$. $\Omega_i$ is the coefficient of the geometric availability of $i$-th site (at full availability $\Omega_i=2\pi$ we estimate that each first order neighbor with the same orientation decreases the solid angle by $-0.39\pi$). $\beta$ is a free variable parameter. A correlation $\beta \sim T^{-1}$ where T is the synthesis temperature, is expected. 
	Since bound fluorine is subjected to migration~\cite{Boukhvalov2016}, we add post-processing consideration of this fact: after bonding of each new fluorine atom, probability of its migration to an unoccupied site in the first surrounding sphere is also calculated in the manner of Eq.~\ref{eq:prob}.

\begin{figure}[!h]
		\begin{center}
			\begin{tabular}{ccc}
				\includegraphics[width=0.27\columnwidth]{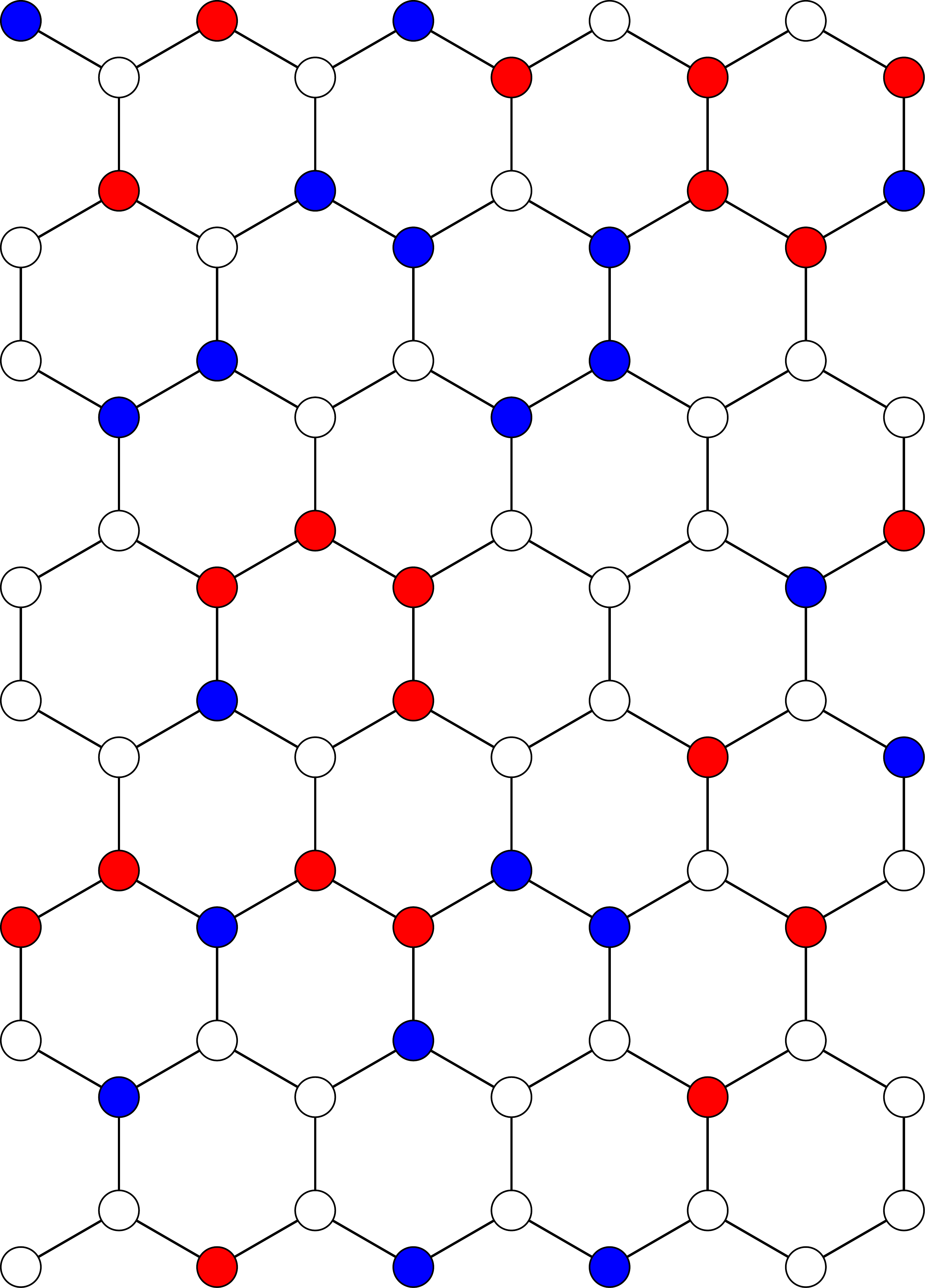} &
				\includegraphics[width=0.27\columnwidth]{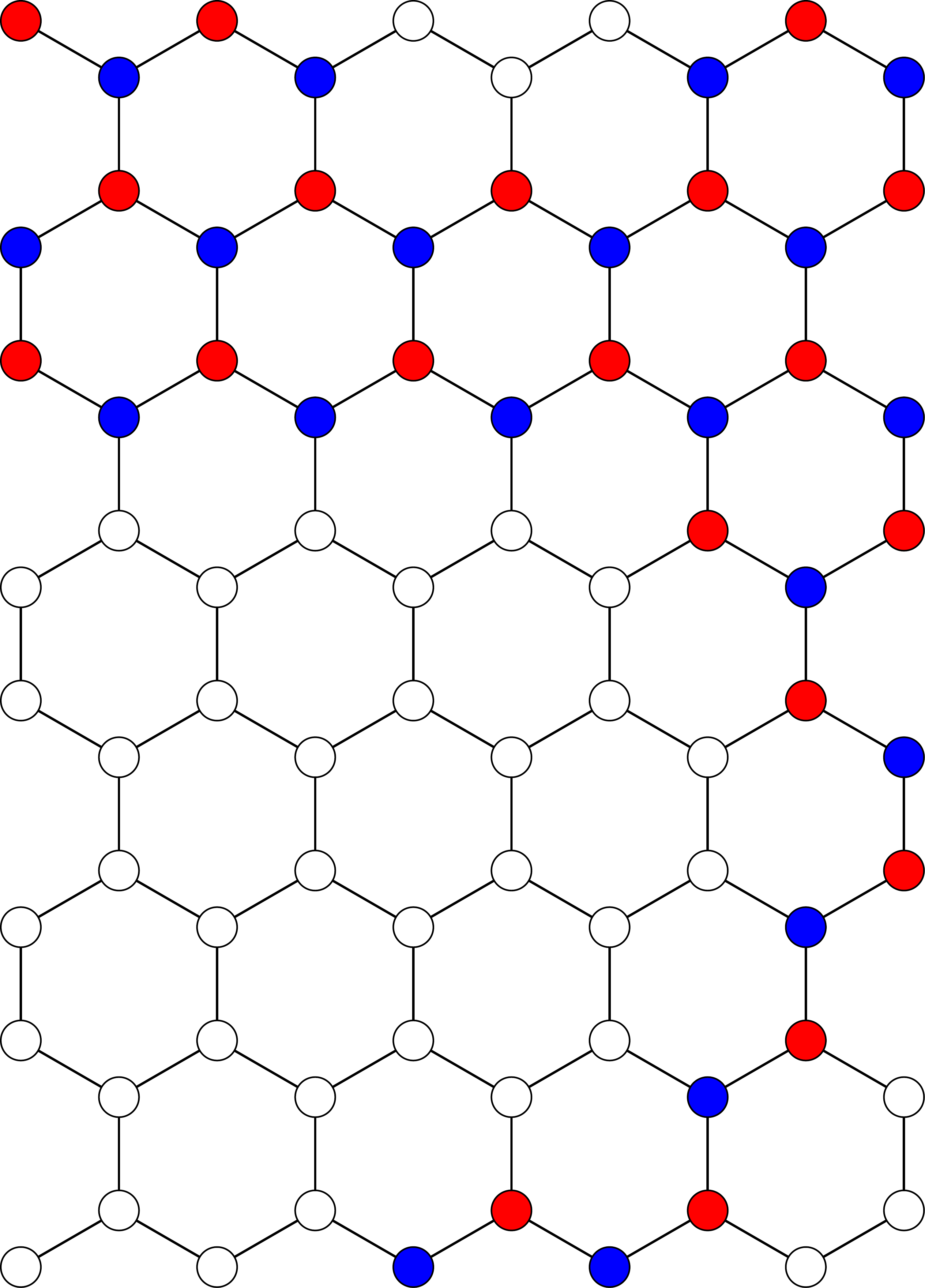}&
				\includegraphics[width=0.27\columnwidth]{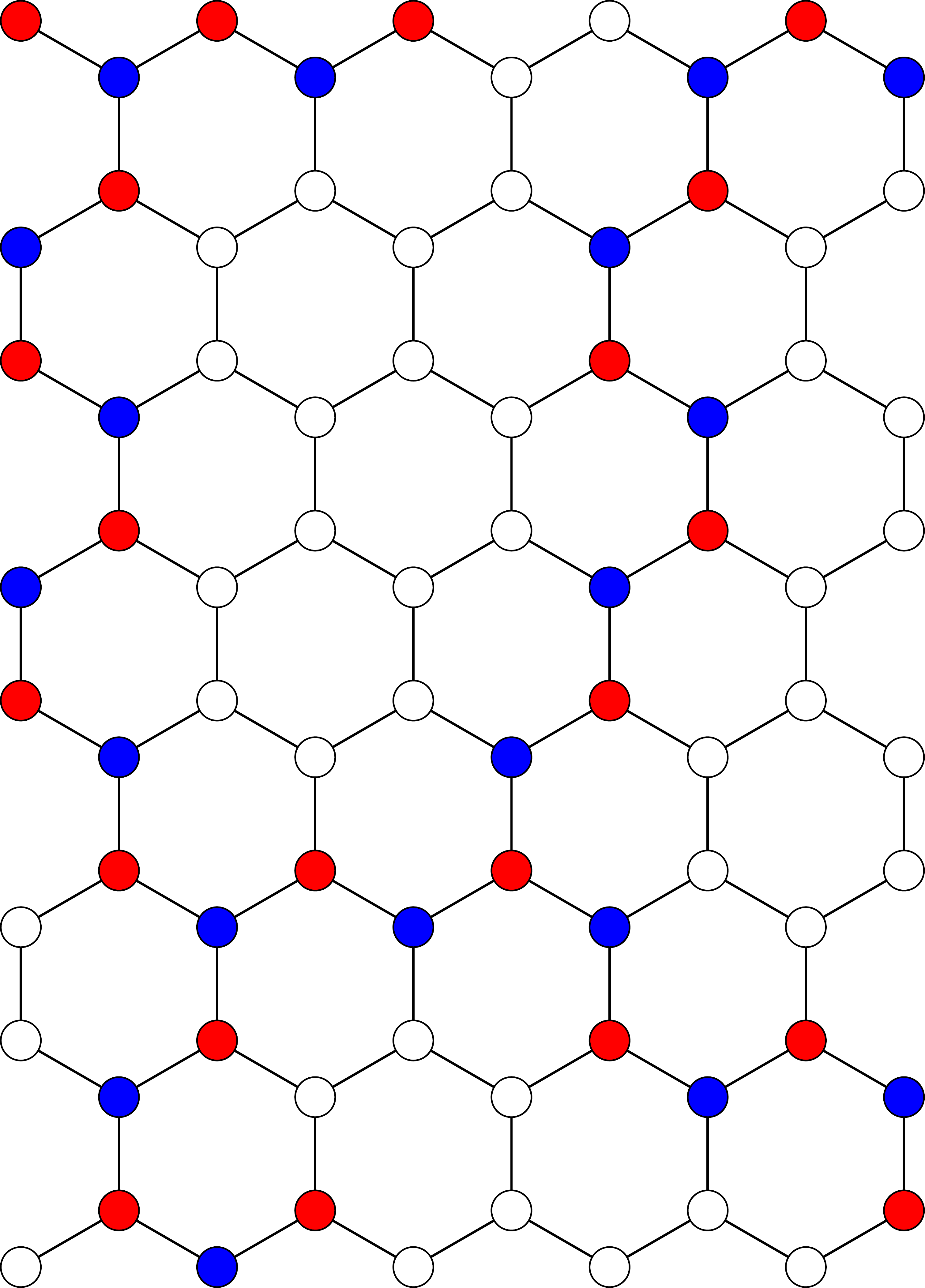}\\
				$\beta=0$ eV$^{-1}$&$\beta=10$ eV$^{-1}$&$\beta=30$ eV$^{-1}$\\
				& & \\
				\includegraphics[width=0.27\columnwidth]{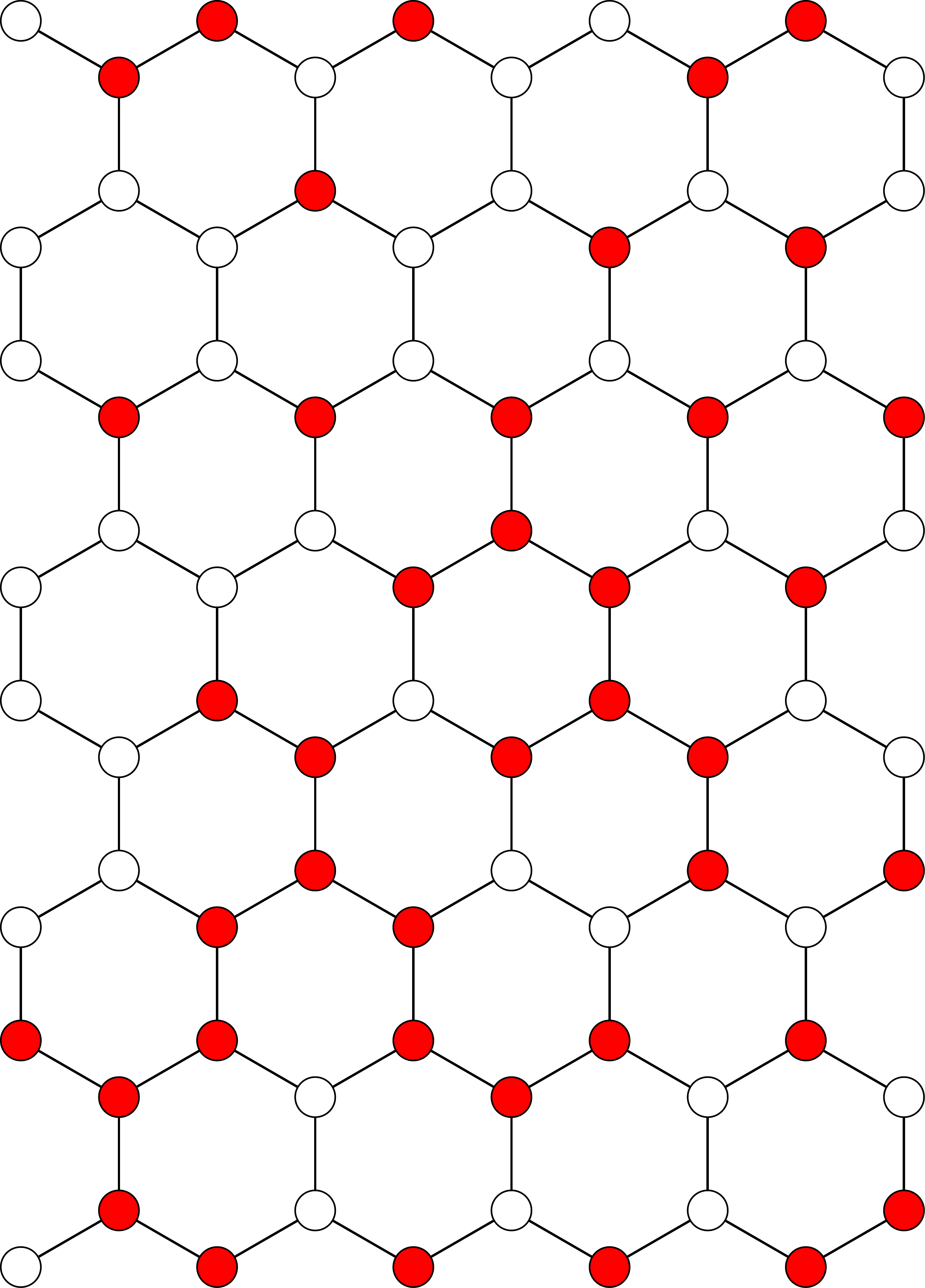} &
				\includegraphics[width=0.27\columnwidth]{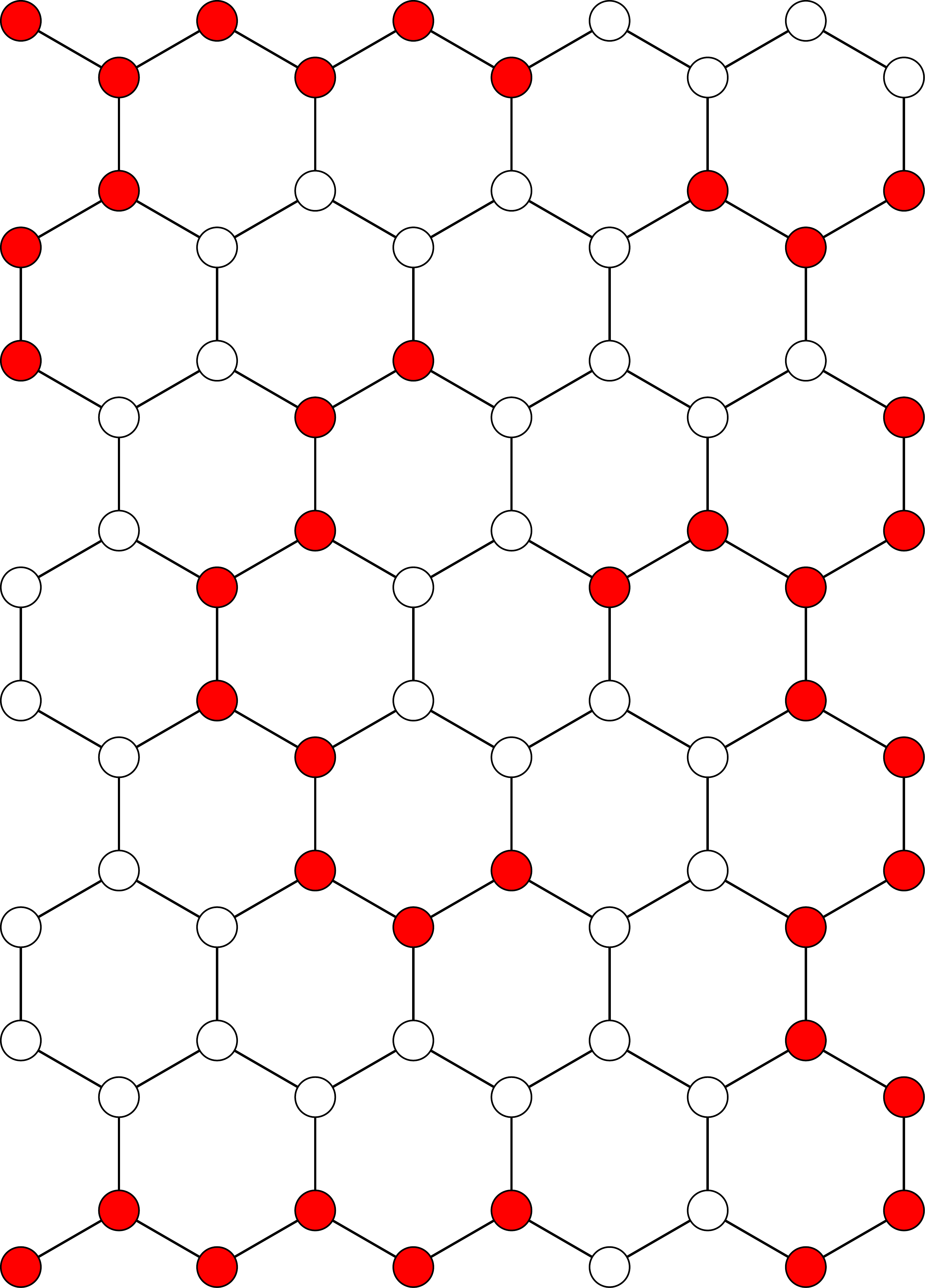}&
				\includegraphics[width=0.27\columnwidth]{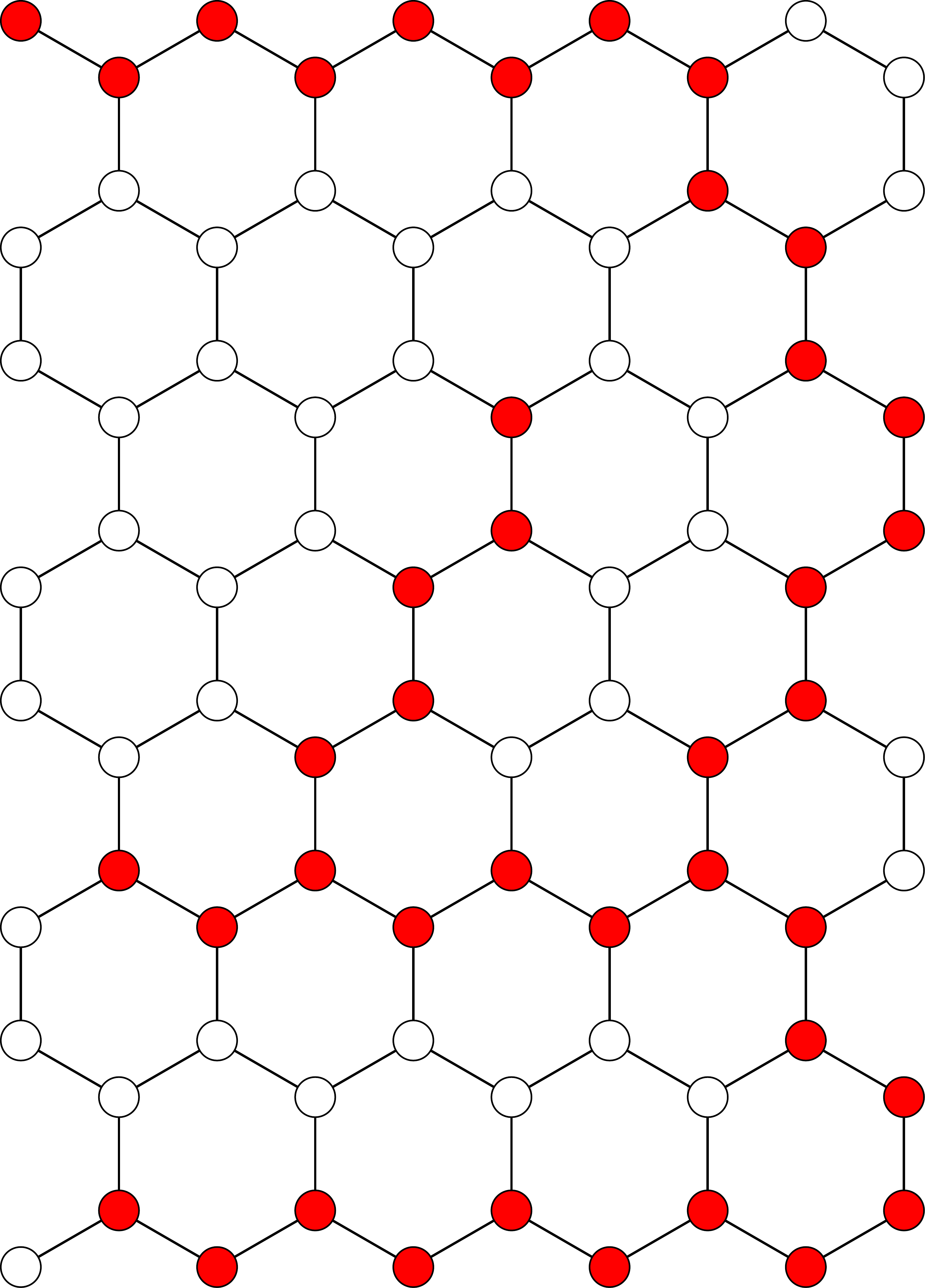}\\
				$\beta=0$ eV$^{-1}$&$\beta=10$ eV$^{-1}$&$\beta=30$ eV$^{-1}$
			\end{tabular}	
		\end{center}
		\caption{CF$_{0.5}$ structures generated at various $\beta$. Top structures correpond to two side fluorination. Bottom - one side fluorination. Random fluorine distribution at $\beta=0$ eV$^{-1}$. Island-like structures at $\beta=10$ eV$^{-1}$ in the case of two side fluorination and formation of pairs, triples and short chains in the case of one side fluorination. Chains formation at $\beta=30$ eV$^{-1}$.}
		\label{fig:c2f_str}
	\end{figure}

	A series of simulations with various $\beta$ and fluorination degree was carried out to verify the adequacy of the above scheme. A set of 5 rectangular C$_{352}$ structures was simulated for each value of $\beta$ and fluorination degree (see Fig.~\ref{fig:cc_ccf},\ref{fig:entropy}) to ensure greater reliability. Examples of obtained results for C$_{80}$F$_{40}$ structures are presented in Fig.~\ref{fig:c2f_str}. We should note, that analysis of the one-side FG is a more complex issue for our model, due to DFT sampling bias towards two-side fluorinated structures. Such bias, as well as the lack of substrate consideration, may lead to underestimation of geometrical stress of one-side fluorinated patterns. Thus we cannot state quantitative patterning results in the case of one-side fluorination. To achieve better results, our model should also be optimized on the DFT dataset of one-side fluorinated structures.
	
	At $\beta=0$ eV$^{-1}$  almost random structures without any distinguishable patterns were obtained. Even if both sides of graphene are accessible for fluorination, several neighboring fluorine atoms can be located on the same side at $\beta=0$, i.e. at extremely high (possible unreachable)  fluorination temperature. The sequentional addition of fluorine to opposite sides of graphene is observed with increasing the value of $\beta$.
	 At $\beta=10$ eV$^{-1}$, fluorine atoms prefer to cluster in island-like structures. Since this behavior is not profitable for one-side fluorination, short fluorine chains are formed in this case. A further increase in $\beta$ leads to the formation of predominantly branched chain structures for both two- and one-side fluorination. Hence, the fluorination proceeds by means of the formation of ribbon-like patterns ~\cite{Boukhvalov2016}. In our approach (Eq.~\ref{eq_E}) there is no difference between armchair and zig-zag chains, while more sophisticated DFT calculations reveal that the structure with alternate zig-zag chains is 0.09~eV more stable than one with armchair chains~\cite{Boukhvalov2016}.

\begin{figure}[!h]
		\begin{center}
			\includegraphics[width=0.60\columnwidth]{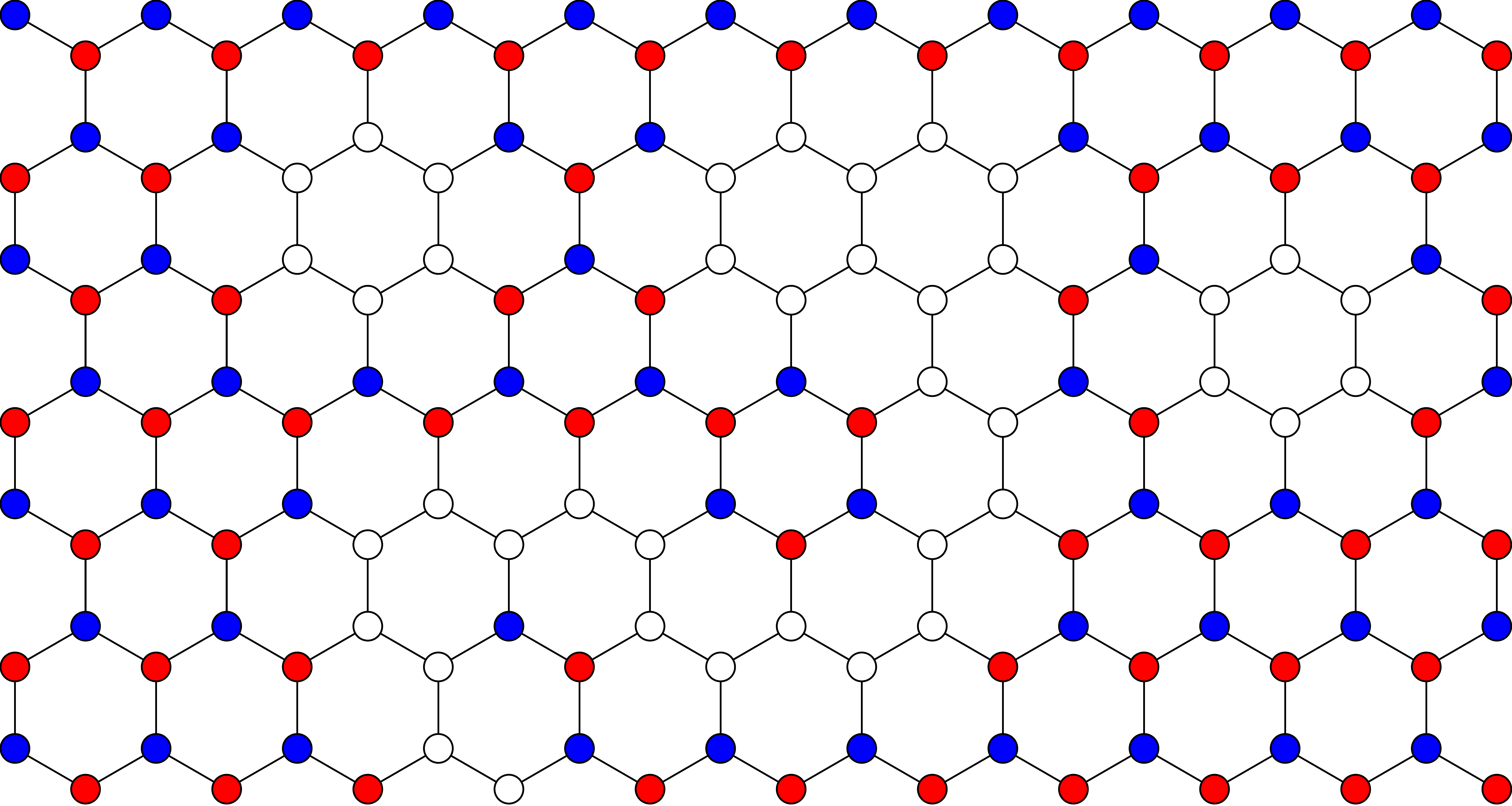}
		\end{center}
		\caption{Benzene, naphthalene units and carbon chains structures in C$_{154}$F$_{114}$ fluorinated graphene, generated at $\beta=30$~eV$^{-1}$}
		\label{fig:c154_str}
	\end{figure}

\begin{figure}[!h]
		\begin{center}
			
			two side fluorination\\
			\includegraphics[width=\columnwidth]{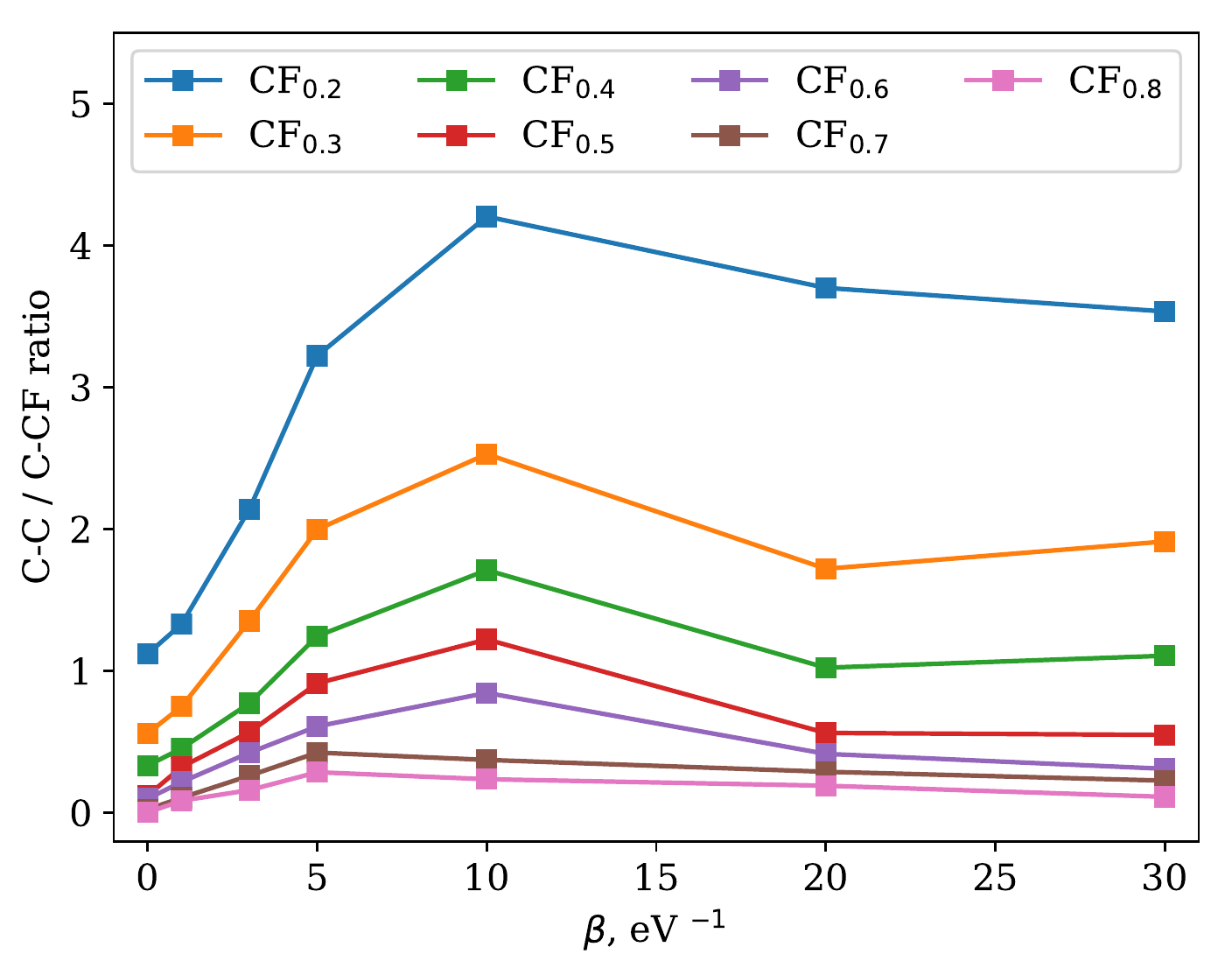}\\
			one side fluorination\\
			\includegraphics[width=\columnwidth]{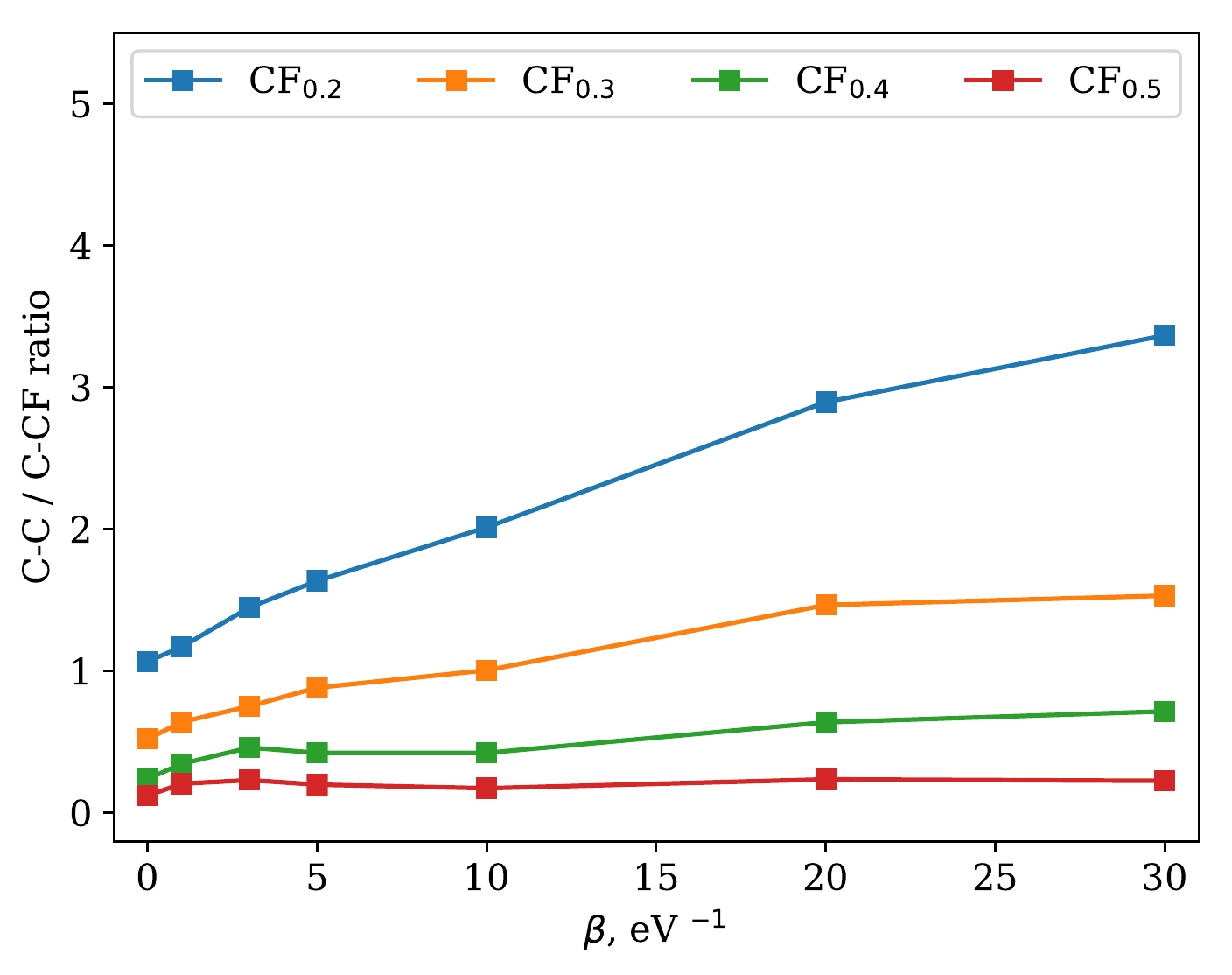}
		\end{center}
		
		\caption{C-C to C-CF atoms type ratio for different fluorine content and fluorination temperature.}
		\label{fig:cc_ccf}
	\end{figure}

\begin{figure}[!h]
		\begin{center}
			two side fluorination\\
			\includegraphics[width=\columnwidth]{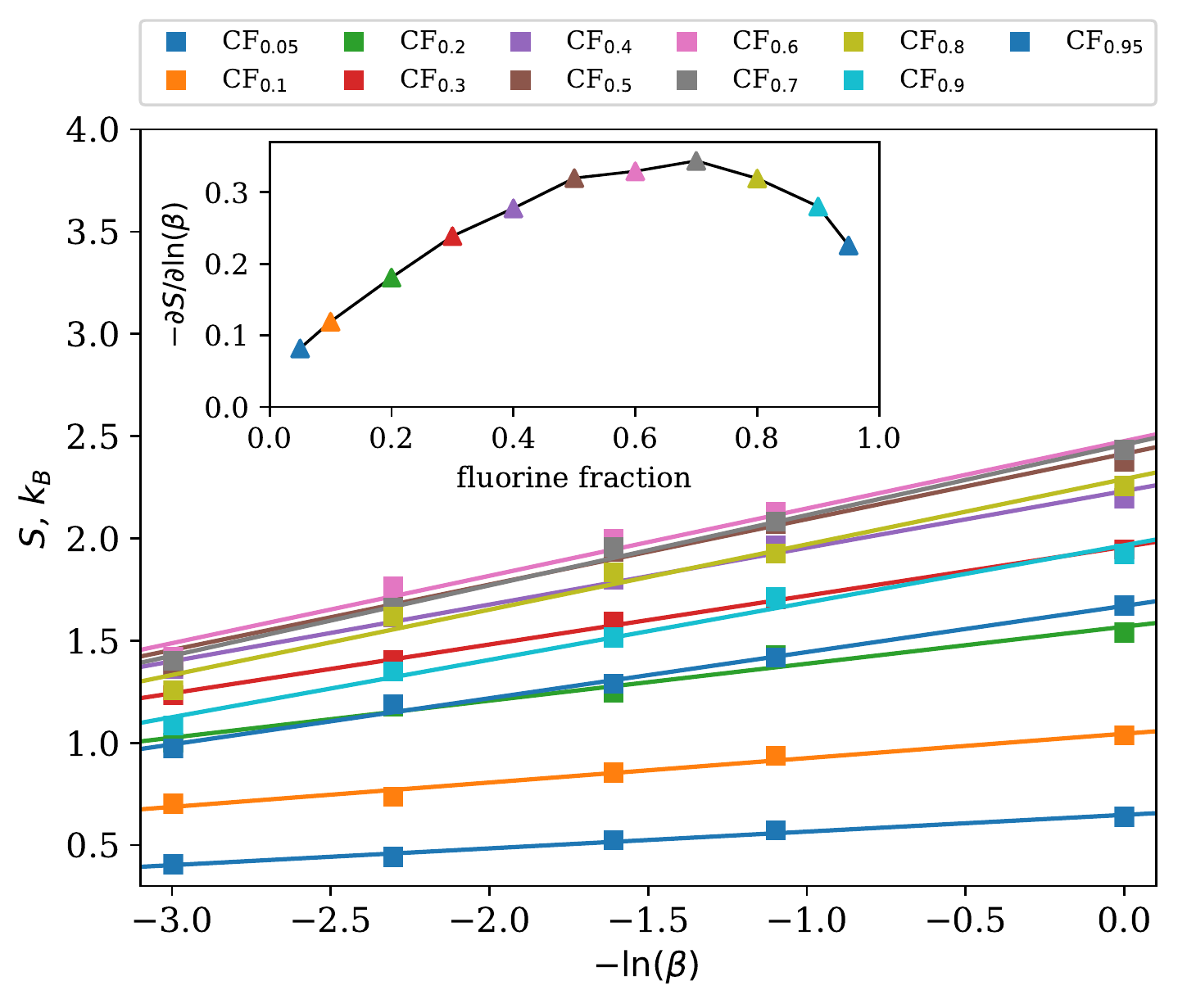}\\
			one side fluorination\\
			\includegraphics[width=\columnwidth]{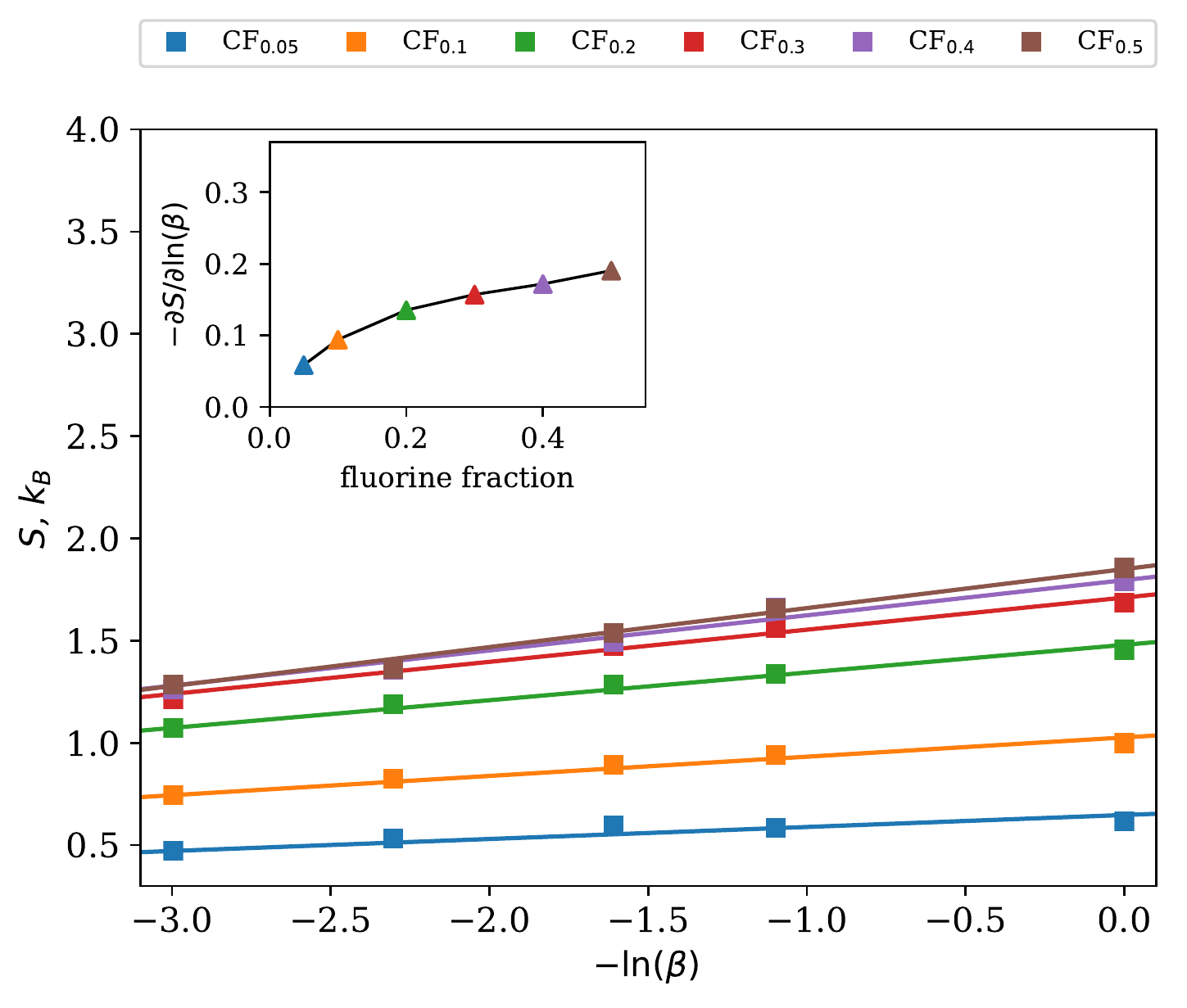}
		\end{center}
		\caption{Structural entropy as a function of $-\ln(\beta)$ for different fluorine fractions. Insert: slope of the linear fit of entropy.}
		\label{fig:entropy}
	\end{figure}

While the formation of one-side FG with more than 50\% of fluorine is thermodynamically unfavorable~\cite{Santos2014}, the properties of the well-known fluorographene CF are widely described~\cite{Nair2010}. In highly fluorinated systems, at high $\beta$, the formation of individual non-fluorinated polyaromatic structures and chains is observed (see Fig.~\ref{fig:c154_str}). Such polyaromatic structures can be considered as short-range isolated graphene quantum dots with widely described prospects and possible applications~\cite{Bacon2014,Nebogatikova2020}.

	To describe obtained structures two metrics were used. The first one is the popular XPS characteristics of fluorocarbons: C-C/C-CF ratio(see Fig.~\ref{fig:cc_ccf}),  where first C-C component is sp$^2$ carbon without neighboring fluorinated C atoms, and C-CF is a bare carbon located next to the CF group. This ratio can be used as a branching metric of the structure: the greater the C-C/C-CF ratio, the less branched fluorine pattern. Another structural metric is structural entropy $S=-\sum \tfrac{N_k}{N} \ln \tfrac{N_k}{N}$ (Fig.~\ref{fig:entropy}), where unique $k$-th substructure (central atom plus first sphere) occurs $N_k$ times out of $N$ (including all symmetrically transformed repetitions). 
	
	The  C-C/C-CF ratio  shows different behavior in case of  one-side and two-side fluorination (Fig.~\ref{fig:cc_ccf}). As described above, in the case of  two-side fluorination the following fluorine patterns are formed as the binding energy increases: scattering; small disordered aggregations of 2-4 atoms; ordered aggregations of larger number of atoms; big islands (predominant). Since the Boltzmann form of Eq.~\ref{eq:prob} leads to higher probability of lower energy structure with $\beta$ increasing, the population of fluorine patterns follows $\beta$ as the binding energy increases. While at low and middle fluorination degree, the greatest fluorine aggregation is achieved at $\beta\sim 10$~eV$^{-1}$, it shifts to  $\beta\sim5$~eV$^{-1}$ with a high fluorination degree due to higher stability of smaller polyaromatic graphene quantum dots~\cite{ribas2011patterning}. The minimum at $\beta\sim 20$~eV$^{-1}$ for CF$_{0.3}$ and CF$_{0.4}$ is associated with a higher branching in relation to $\beta\sim30$~eV$^{-1}$.  Unlike the two-side fluorination, with one-side there are no explicit maxima at $\beta\sim 10$~eV$^{-1}$, which is associated with an absence of large aggregated clusters. The formation and further transition of small aggregates into short chains are responsible for a slight excess at $\beta\sim 3$~eV$^{-1}$. Subsequent $\beta$ growth leads to an increase in the length of chains and a decrease in their branching.
	
	As expected, structural entropy linearizes with the $\beta$ on a logarithmic scale (Fig.~\ref{fig:entropy}). In two-side fluorinated structures entropy is generally slightly higher at $-\ln\beta\sim-3$ with noticeably faster growth (see insert in Fig.~\ref{fig:entropy}), and it reaches saturation at higher fluorine content, due to $\times\tfrac{3}{2}$ times increase of possible states for each carbon. Based on thermodynamics analogy: $ - \left. \frac{\partial S}{\partial \ln\beta} \right|_{V} = c_v$, higher value of $c_v$ leads to a greater susceptibility of the system to $\beta$ changes. Following these considerations further, we expect the highest structural variability for CF$_{0.7}$ and CF$_{0.5}$ for the two- and one-side fluorination, respectively. 

    We compared our results with several experimental studies for both one and two side fluorination cases. Such a comparison is more qualitative than quantitative, due to the small and noise statistics of microscopy and the data ambiguity from other methods. In Ref.~\cite{Jeon2011} free-standing graphene was fluorinated by XeF$_2$ for 1-5 days at 250~$^\circ$C. According  to the transmission electron micrographs of obtained FG, it consists of highly fluorinated domains, surrounding planar, graphene-like domains. Based on Fig.2a of Ref.~\cite{Jeon2011} we can add, that characteristic size of heterogeneity is about 1-2 nm. Another sample of FG was obtained by treating the intercalate    C$_{8}$Br with a BrF$_3$/Br$_2$ solution for 30 days at room temperature. Based on atomic force    microscopy, authors of Ref.~\cite{Asanov2013} claim that there are $\sim2$~nm regions as well as 0.5 nm round-shape areas of non-fluorinated carbon on their FG CF$_{0.4}$. Similar results were obtained in NMR study~\cite{Vyalikh2013}. In above works, we could not find any information that clearly indicates the value of $\beta$ for our model. Based on synthesis temperature, we can estimate $\beta=22$~eV$^{-1}$ for XeF$_2$ and $\beta=38$~eV$^{-1}$ for BrF$_3$/Br$_2$. According to our results, in both cases brunched chains should predominantly be formed. The main difference between $\beta=22$~eV$^{-1}$ and $\beta=38$~eV$^{-1}$ is the degree of branching: lower $\beta$ leads to higher disorder and stronger fluorine aggregation, which results in a smaller characteristic size of the pattern. 
    
   As mentioned above, we cannot state quantitative patterning results in the case of one-side fluorination. Keeping this in mind, we analyze the experimental data on one-side fluorination with plasma (SF$_6$). The functionalization was performed with microwave induced SF$_6$ plasma, with ions acceleration at 0 (exposure 20 min) and 1 kV (exposure 10 min)~\cite{Struzzi2017}.  Graphene/Cu kept more than 20~cm away from the source to minimize surface damage. Based on C 1s XPS decomposition of FG we reconstruct the corresponding $\beta$ (table~\ref{xps results}). As expected, high voltage plasma generates a lower $\beta$ structure. The resulting structure mainly consists of short chains (2-4 atoms for $\beta=7$~eV$^{-1}$ and 2-6 atoms for $\beta=13$~eV$^{-1}$ at given concentrations and $\beta$.

	\begin{table}[!h]
		\caption{XPS results of Ref.~\cite{Struzzi2017} and calculated $\beta$ for the one-side fluorination case. }
		\begin{center}
			\begin{tabular}{|c|c|c|c|}
				\hline
				plasma & structure & C-C/C-CF & $\beta$, eV$^{-1}$\\
				\hline
				SF$_6$, 0 kV & CF$_{0.10}$ & 13.8 & 13\\
				\hline
				SF$_6$, 1 kV & CF$_{0.08}$ & 4.9 & 7\\
				\hline
			\end{tabular}
			\label{xps results}
		\end{center}

	\end{table}
\section*{Conclusions}
    In summary, the new method of modeling FG system is presented. Being optimized on a large DFT dataset, it gives a good agreement with the known theoretical and experimental results. We found that varying the synthesis temperature can lead to various structural features: from short chains for low $\beta=1/k_B T$ (reachable only in plasma synthesis), through bulk or longer chains at $\beta\sim10$~eV$^{-1}$ (T$\sim500~^\circ$C) for one side and two-side fluorination, correspondingly, to chains elongations and brunching decrease at higher $\beta$. Another prediction is the growth of graphene quantum dot's size with the synthesis temperature increase. The accuracy of our model can be estimated by varying the temperature with the same synthesis method. The best region to our prediction examination is compounds with the highest structural variability CF$_{0.5-0.8}$ for the two-side FG and  CF$_{0.4-0.5}$ for the one-side FG. We believe that our work can help in further studies of FG based materials.
	
	Source code and ready-to-use build of described structure generator \underline{\href{https://github.com/yamarus/genCF/}{is available}} on GitHub~\cite{genCF}.
\section*{Acknowledgments}
The reported study was funded by RFBR, project number 19-32-60012.
	\bibliographystyle{ieeetr}
	\bibliography{biblio}
\end{document}